\documentclass[a4paper,fleqn]{cas-dc}

\usepackage[numbers]{natbib}
\usepackage{tabularx}
\usepackage{multirow}  
\usepackage{float}
\usepackage{xcolor}
\usepackage{enumitem}
\usepackage{float}

\hypersetup{
    colorlinks=true,
    linkcolor=blue,
    citecolor=blue,
    urlcolor=blue
}

\def\tsc#1{\csdef{#1}{\textsc{\lowercase{#1}}\xspace}}
\tsc{WGM}
\tsc{QE}
\tsc{EP}
\tsc{PMS}
\tsc{BEC}
\tsc{DE}

\begin{document}
\let\WriteBookmarks\relax
\def\floatpagepagefraction{1}
\def\textpagefraction{.001}
\shorttitle{ }

\title [mode = title]{pAE: An Efficient Autoencoder Architecture for Modeling the Lateral Geniculate Nucleus by Integrating Feedforward and Feedback Streams in Human Visual System}

\author{\textcolor{black}{Moslem Gorji}}

\author{\textcolor{black}{Amin Ranjbar}}
\cormark[1]
\ead{amin.ranjbar1988@gmail.com}

\author{\textcolor{black}{Mohammad Bagher Menhaj}}

\affiliation{organization={Department of Electrical Engineering, Amirkabir University of Technology},
                city={Tehran},
                country={Iran}}

\cortext[cor1]{Corresponding author}

\begin{abstract}
The visual cortex is a vital part of the brain, responsible for hierarchically identifying objects. Understanding the role of the lateral geniculate nucleus (LGN) as a prior region of the visual cortex is crucial when processing visual information in both bottom-up and top-down pathways. When visual stimuli reach the retina, they are transmitted to the LGN area for initial processing before being sent to the visual cortex for further processing. In this study, we introduce a deep convolutional model that closely approximates human visual information processing. We aim to approximate the function for the LGN area using a trained shallow convolutional model which is designed based on a pruned autoencoder (pAE) architecture. The pAE model attempts to integrate feed forward and feedback streams from/to the V1 area into the problem. This modeling framework encompasses both temporal and non-temporal data feeding modes of the visual stimuli dataset containing natural images captured by a fixed camera in consecutive frames, featuring two categories: images with animals (in motion), and images without animals. Subsequently, we compare the results of our proposed deep-tuned model with wavelet filter bank methods employing Gabor and biorthogonal wavelet functions. Our experiments reveal that the proposed method based on the deep-tuned model not only achieves results with high similarity in comparison with human benchmarks but also performs significantly better than other models. The pAE model achieves the final $99.26\%$ prediction performance and demonstrates a notable improvement of around $28\%$ over human results in the temporal mode.
\end{abstract}


\maketitle

\section{Introduction}

Understanding the complex functioning of the visual pathways poses a significant challenge within the computational neuroscience area. The study of the human visual pathway has the potential to improve various scientific disciplines, contributing to a diverse range of outcomes including the treatment of vision-related disorders like amblyopia \cite{duffy2023human} and also developing the brain-computer interfaces (BCI) systems utilized in neurorehabilitation \cite{vavoulis2023review}, gaming and virtual reality \cite{hadjiaros2023virtual}, and also eye-related communications \cite{vortmann2022multimodal}. Many researchers have attempted to simulate the human brain, focusing more on modeling the vision system. This includes creating models for the ventral pathway \cite{riesenhuber1999hierarchical, masquelier2007unsupervised, zhuang2021unsupervised}, the lateral geniculate nucleus (LGN) \cite{einevoll2022lateral, mounier2021deep}, combinations of the LGN and the primary visual cortex (V1) \cite{ji2021retina}, and also the ventral pathway \cite{zabbah2014impact, davoodi2023classification}.

Visual information processing starts with the retina, LGN, and V1, through the visual pathway where it is analyzed before being further processed in higher visual regions. The received input images at the retina will be subjected to higher-level cognitive processing for extracting abstract information at late areas, specifically from V2 to the inferotemporal cortex (IT) and the middle temporal (MT) areas \cite{kruger2012deep}. The LGN \cite{muller2021mapping} serves as a primary subcortical processing stage in the human visual pathway playing a significant role in preparing data to be transmitted to later areas for crucial tasks such as decision-making \cite{davoodi2023classification}, perception, and motion detection \cite{rokszin2010visual}. In computational modeling of the LGN area, a common approach involves simulating patterns in LGN responses through filtering methods such as the Difference of Gaussians (DoG) and Gabor filter bank \cite{azzopardi2012corf, davoodi2023classification}. Deep convolutional neural networks (CNNs) have also been frequently employed to predict LGN responses, incorporating both the history of neuronal firing and the spatiotemporal characteristics of visual inputs \cite{mounier2021deep}.

Besides, when the input image is transmitted to the primary visual cortex, a portion of LGN calculations will be connected to the V1 area. This area uses a feedforward pathway in which spikes travel from the LGN to layer 6 of V1 and are subsequently returned to the output layer of the LGN through a feedback pathway. Within V1 layer 6, there are three distinct channels: channel 1 receives spikes from LGN layers 3/4, channel 2 from LGN layers 5/6, and channel 3 from LGN layer K. Each channel transmits the processed spikes to the corresponding output layer in the LGN. The synaptic connections between the output layer of the LGN and V1 layer 6 consist of one-to-one excitatory synapses \cite{ji2021retina}. Importantly, when modeling the LGN area, it is crucial to consider that while a brain structure may preserve a substantial portion of the information processed by the retina, the average firing rate of the LGN is significantly lower than the retinal neurons within this feedforward-feedback structure. For every ten spikes that the retina delivers, only four are fired by the LGN. Upon departing from the retina, the visual data flow is significantly reduced by the time it arrives at the V1. This reveals an information loss through visual processing which has been well-established in neuroscientific research \cite{zhong2021neural, uglesich2009stimulus}.

\begin{figure}
	\centering
	\includegraphics[width=1\columnwidth]{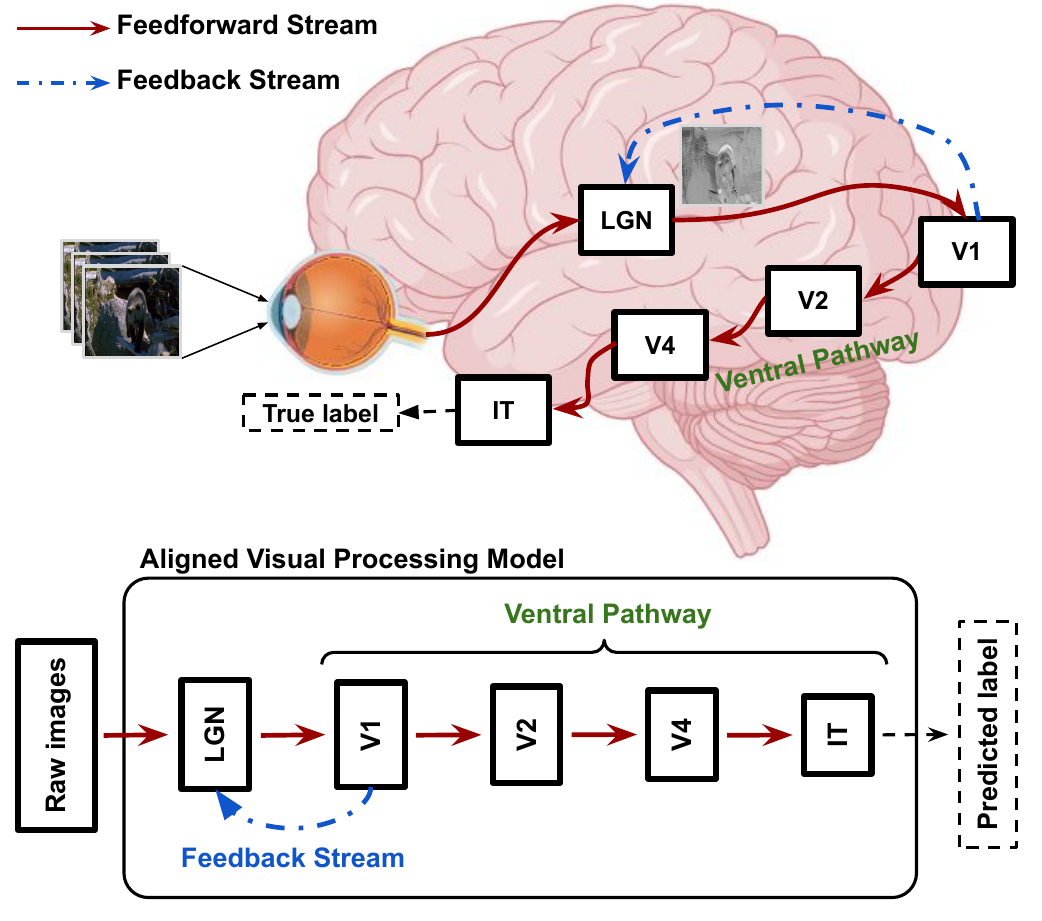}
	\caption{The schematic of both biological and artificial visual processing systems in humans. The visual information processing initiates in the LGN area where the input data is crucially prepared for further processing through the ventral pathway in late visual areas, specifically from V2 to the IT. The developed model aims to mimic the biological system and predicts labels that closely resemble the true labels.}
	\label{FIG:1}
\end{figure}

The visual system has been a major factor in recent advances in computer vision and neuroscience. Numerous structural characteristics are shared by the ventral route of the visual cortex and the CNN architecture, which are frequently utilized in computer vision \cite{zhong2021neural}. Moreover, deep artificial neural networks featuring spatially repeated processing have emerged as the most effective models for simulating the ventral visual processing stream in primates \cite{kubilius2018cornet} as depicted in \hyperref[FIG:1]{Fig. \ref{FIG:1}}. Computer vision moved into a deep learning paradigm with the introduction of AlexNet, but it also introduced a novel method of "system identification" and an approximation of how neurons in the visual cortex construct robust and invariant representations of objects \cite{linsley2024performance}. A highly debated question in the field is whether and how the process of image reconstruction from the V1 to the LGN can enhance decision-making within the visual cortex. To date, there have been very few empirically published studies on image reconstruction from the V1 to LGN in the field of computational neuroscience. Additionally, the experimental investigation of image reconstruction within the visual pathway using various models has not yet been explored.
In the current study, we primarily focused on modeling the LGN and simulating the ventral pathway using hierarchical classification models. A key aspect of the first step in our investigation was the feedback mechanism model from the V1 to the LGN, which is vital to consider for developing a biologically plausible visual processing pipeline. We examined multiple approaches to optimize the LGN-V1 pathway, aiming to enhance both the accuracy and efficiency of our models. Therefore, we propose a pruned AlexNet autoencoder (pAE) model to simulate various functions including feedforward and feedback interactions within the LGN area. This customized model employs a single-layer convolutional encoder-decoder to approximate both the forward and backward flow of information between the LGN and V1 areas. We examined the performance of this model using both pre-trained and trained versions to replicate outputs of LGN, which are built on the CNN architecture. Also, to identify the most effective approach and conduct a comparative analysis, we designed a range of filtering models utilizing Gabor Filter Banks (GFB), and Discrete Wavelet Transform (DWT) to accurately model the LGN region. For efficient GFB, we applied a tailored feature extraction method by controlling the number of low-frequency feature maps for the reconstruction phase. These filtering models allow for a more comprehensive analysis of neural pathways involved in visual perception. In the next phase, we developed a convolutional model to accurately predict the true labels of images generated by the LGN for two categories: animals and non-animals. Following our analysis, we proceed to evaluate the outcomes with the widely recognized HMAX model and also a real psychophysical experiment that includes 30 participants.

\section{Materials and Methods}
\subsection{Experimental dataset}
\begin{figure}
	\centering
	\includegraphics[width=1\columnwidth]{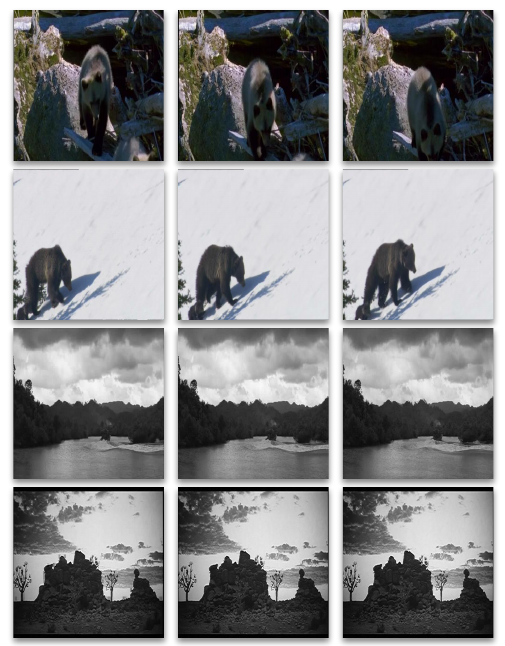}
	\caption{Captured image frames depicting animal and non-animal movements. The upper two rows display two series of three consecutive frames, each capturing an animal in motion. In contrast, the lower two rows showcase two sets of three consecutive frames capturing movements of non-animal objects such as leaves, sea, waterfall, humans, etc.}
	\label{FIG:2}
\end{figure}

We utilized a collection of videos from nature documentaries mentioned in the paper \cite{zabbah2014impact}. For every video, we examined three consecutive frames where the camera was stationary and captured a barely identifiable object - it could be either an animal or a non-animal - in motion. The slight movements of the object across successive frames aided in locating the concealed object, thereby facilitating classification. This dataset comprises 102 animal images and 102 non-animal images, each comprising a sequence of three frames extracted from the input video. The animal images cover various categories including mammals, insects, fish, reptiles, and birds. All collected frames are grayscale, resized to 256 x 256 pixels (as illustrated in \hyperref[FIG:2]{Fig. \ref{FIG:2}}), and have a frame rate of 13 milliseconds per frame to enhance motion detection in the videos.

\subsection{Modeling framework for the biological visual processing}
The biological visual processing system receives visual stimuli from the retina, a layer of light-sensitive tissue, and then converts them into electrical impulses. These impulses are first sent to the LGN for initial processing and then transmitted to the V1 for further processing. The output from V1 then serves as input for subsequent layers within the visual system. Through our modeling framework, we utilized two distinct image-feeding forms: the temporal and non-temporal modes. In the non-temporal mode, the input image is transferred directly from the retina to the visual cortex, bypassing the LGN model. In the temporal or LGN-enabled mode, the image is first processed by the LGN model before being transferred to the visual cortex for further processing \cite{zabbah2014impact}.

\begin{figure}[htbp]
	\centering
	\includegraphics[width=1\columnwidth]{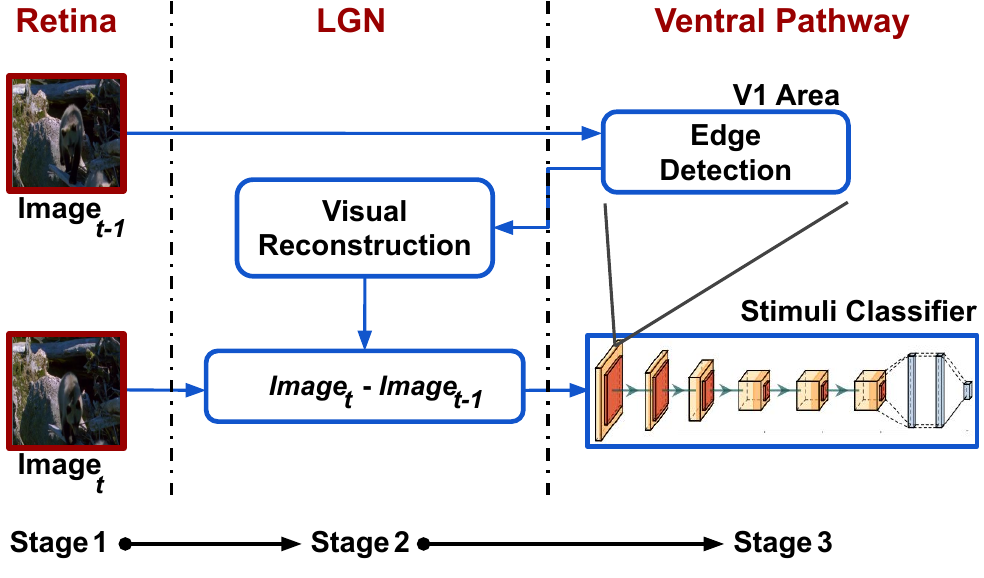}
	\caption{Proposed modeling framework for processing visual data in LGN and V1 areas. This model demonstrates the processing of the initial image by the LGN area for edge detection in V1. The image is then reconstructed for further computation of difference with the second consecutive frame for image classification, simulating the ventral visual cortex.}
	\label{FIG:3}
\end{figure}

As depicted in \hyperref[FIG:3]{Fig. \ref{FIG:3}}, the proposed model for simulating the LGN-V1 interactions comprises three following stages:

\textit{Stage 1:} It corresponds to the retina where it receives the current and subsequent image frames to transmit to the LGN layer for edge detection tasks in V1. It also computes the movements using the difference of consecutive images.

\begin{flalign}
 E[f(x,y;t-1)] &= \left |\bigtriangledown f(x,y;t-1)  \right | \nonumber\\
 &\approx \left |\bigtriangledown f_{x}  \right | + \left |\bigtriangledown f_{y}  \right| \label{eq1}
\end{flalign}

The input image is denoted as $f(x,y;t-1)$ in \hyperref[eq1]{Eq. \ref{eq1}}, and $E[f(x,y;t-1)]$ represents the edge detection operation performed on the image.

\textit{Stage 2:} It simulates the functions that happen in the LGN area where the original image is reconstructed from the image of edges detected in V1. The symbol $R[.]$ in \hyperref[eq2]{Eq. \ref{eq2}} signifies the reconstruction process occurring in LGN-V1 interactions which results in $f(x,y;t-1)$, called the reconstructed image. 
\begin{flalign}
     R[E[f(x,y;t-1)]] &= \hat{f}(x,y;t-1) \nonumber\\ &= \iint_{}^{} \bigtriangledown f(x,y;t-1) dx dy \label{eq2}
\end{flalign}

This stage also calculates the difference between the second input frame and the reconstructed image to approximate changes for higher-level visual tasks like the classification of image objects. Please note that the feedback mechanism leads to the calculation of the difference between consecutive images due to a phase-reversed influence pattern where the off-center cell is excited and on-center geniculate cells are inhibited through a V1 on-center cell \cite{wang2006functional}. This pattern shows that LGN receptive fields aligned over simple cell zones of the same polarity decrease their responses, while those of opposite polarity increase their responses with feedback. This phase-reversed influence pattern, observed in recent neurophysiology studies, supports the idea that feedback describes the general coding theory in the brain \cite{alitto2003corticothalamic, callaway1998local}. As depicted in \hyperref[FIG:3]{Fig. \ref{FIG:3}}, the function $D[.]$ in \hyperref[eq3]{Eq. \ref{eq3}} represents the difference operation between $f(x, y;t)$ and $\hat{f}(x,y;t-1)$ to generate the LGN output for subsequent regions.
\begin{eqnarray}
     D[f(x,y;t), \hat{f}(x,y;t-1)] = f(x,y;t) - \hat{f}(x,y;t-1) \label{eq3}
\end{eqnarray}

\textit{Stage 3:} All processing that happens in the ventral visual pathway is modeled as a hierarchical classifier to forecast the input image labels. Through this stage, we utilize the AlexNet and HMAX (Hierarchical MAX) models as general classifiers. Each internal layer of these models corresponds to the specific biological visual region, allowing for a comprehensive analysis and classification of the input data.

\subsection{Investigation into LGN-V1 information flow}
The LGN and V1 regions are crucial in the early phases of visual processing. The flow of information between these regions involves complex feedforward and feedback mechanisms that are vital for processes such as edge detection and visual reconstruction in the brain. Through the feedforward pathway, the retina captures primary visual stimuli, which are then processed and transmitted to the LGN by retinal ganglion cells. The LGN acts as a relay station, organizing visual information before being sent to V1. As well, the feedback pathway includes transmitting information from V1 to the LGN, which is critical for exact data processing and enhancing the perception of visual stimuli \cite{callaway2004feedforward, alitto2003corticothalamic, angelucci2006contribution}.

To model the LGN-V1 information flow, we incorporated four distinct models into our proposed visual processing framework: Discrete Wavelet Transform (DWT), and a proposed pruned autoencoder model. We explored the feedforward-feedback mechanism where the LGN-processed images are transmitted to the V1 area and then loop back to the LGN for precise perception.

\subsubsection{CNN model as an approximator for LGN-V1 connection}
\begin{figure}[htbp]
	\centering
	\includegraphics[width=1\columnwidth]{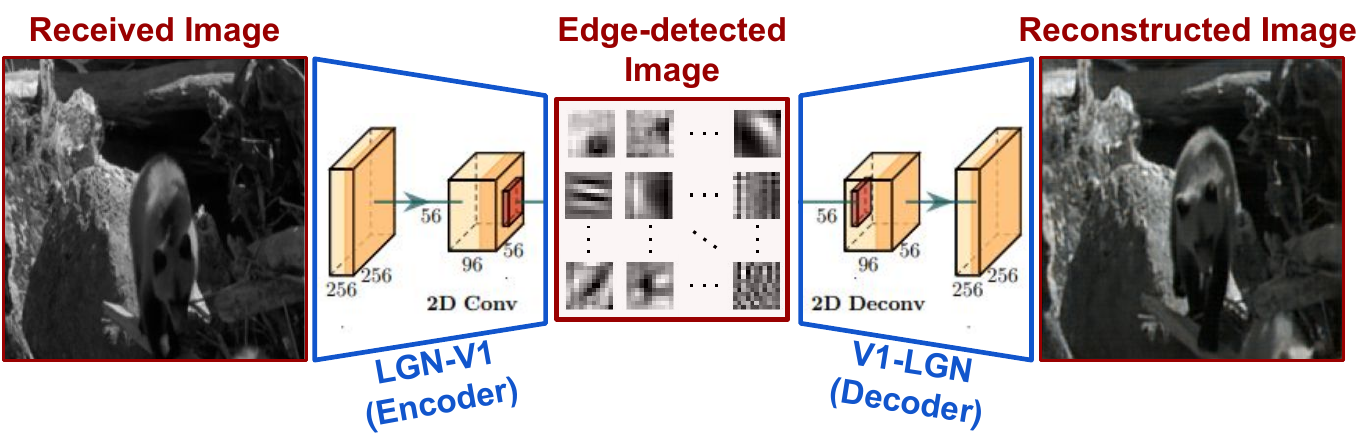}
	\caption{The autoencoder model approximates the LGN-V1 connection using the pAE model. This model builds a more simplified architecture that mimics the actions of the visual cortex's first layer, or V1. Using a single convolutional layer as an encoder and a deconvolutional layer that functions as a decoder, this model effectively captures the characteristics of the first visual layer and also simulates the backward path from V1 to the LGN. The image reconstruction from the edge-detected outputs based on terminal convolutional kernels ensures an accurate representation of the LGN region. Overall, the innovative design of pAE demonstrates a biologically plausible model for visual processing, enhancing our understanding of the LGN-V1 interactions through sophisticated simulation techniques.}
	\label{FIG:4}
\end{figure}

AlexNet is a well-known model in the neuroscience community due to its compact architecture and high similarity to the natural visual system of the human brain \cite{lindsay2021convolutional, jozwik2018deep, wen2018neural}. AlexNet consists of eight trainable layers, the first five layers are convolutional and the remaining three layers are weighted fully connected layers. This structure allows it to effectively simulate the hierarchical processing observed in LGN-V1 connectivity \cite{jozwik2018deep}. 
In this section, we designed an autoencoder model based on a pruned version of the AlexNet model which will be called pAE going forward in this research. Specifically, this model represents a more shallow structure that closely resembles the behavior of the initial layer V1, resulting in more biologically plausible modeling. As depicted in \hyperref[FIG:4]{Fig. \ref{FIG:4}}, to establish an aligned artificial model, pAE uses one convolutional layer as a single-layer encoder, which mimics the characteristics of the first visual layer and provides the highest level of simulation accuracy. Also, the pAE comprises a deconvolutional layer used to simulate a single decoder layer for the backward path from V1 to the LGN, reconstructing the image from the edge-detected outputs for the LGN section as comprehensively described in stage 2.

\subsubsection{DWT model as an approximator for LGN-V1 connection}
\sloppy
Time-frequency analysis using DWT has gained widespread adaptation in computational neuroscience for analyzing time series data \cite{chen2017high, eskikurt2024evaluation}. It serves as a crucial tool in signal processing, allowing the decomposition of a given signal into multiple sets, with each set representing a time series of coefficients that describe the signal's time evolution within a specific frequency band \cite{percival2012wavelet}. Here, we use DWT offering multi-resolution analysis of the image to decompose input images into different frequency components at various scales. This is achieved through the application of wavelet filters, which can capture both spatial and frequency information effectively. This ability provides us with both time and frequency localization and makes it particularly suitable for image-processing tasks such as edge detection \cite{divakar2022image}. Thus, \hyperref[eq4]{Eq. \ref{eq4}} represents the wavelet decomposition on a given 2D image $f(x,y)$ which involves convolving the input signal with the biorthogonal filters.
\begin{flalign}
    &E[f(x,y; t-1)] = W_{\psi}(j, m, n) \nonumber\\ &= \sum_{x} \sum_{y} f(x, y; t-1) \psi_{j, m, n}(x, y; t-1) \label{eq4}
\end{flalign}

where $W_{\psi}$ denotes the detail coefficients, j represents the scale and m,n indicates dilation parameters. Also, the 2D biorthogonal wavelet function is derived from the utilization of a wavelet pair, and it generally has the following decoupled form for decomposition:
\begin{flalign}
    \psi(x, y) = \psi(x) \psi(y) \label{eq5}
\end{flalign}

\begin{figure}
	\centering
	\includegraphics[width=1\columnwidth]{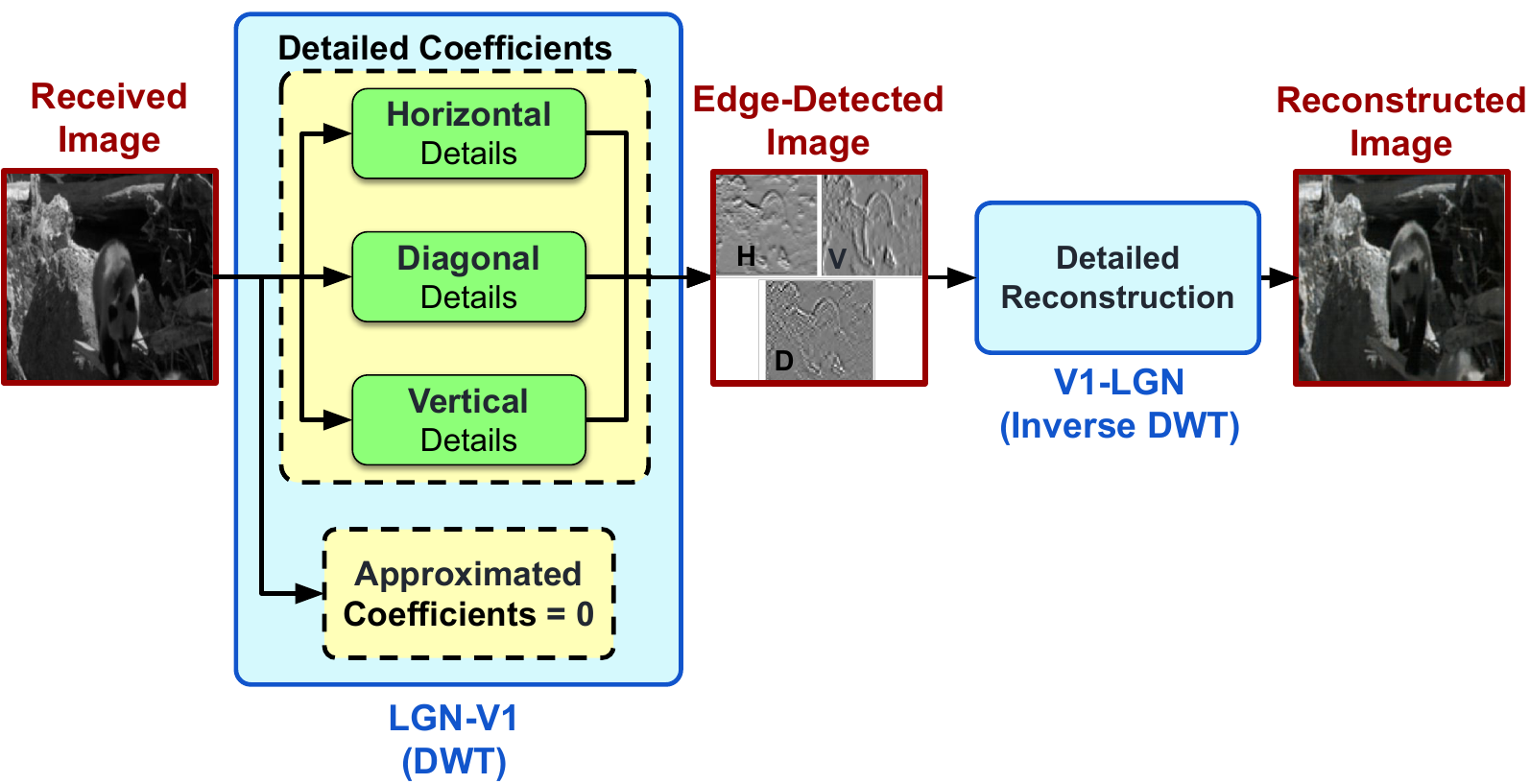}
	\caption{The multi-resolution analysis to approximate the LGN-V1 connection using DWT. This model decomposes input images into different frequency components at various scales. The proposed model uses detailed wavelet coefficients to generate edge-detected images at V1 and consequently applies inverse DWT on these coefficients to reconstruct the images in the backward stream.}
	\label{FIG:5}
\end{figure}

By \hyperref[FIG:5]{Fig. \ref{FIG:5}}, the proposed model mimics the function of V1 in biological systems for the image transformation from the retina to the V1, which attempts to preserve as much as possible all the details of the image for further processing in the ventral pathway. We utilize Gabor and biorthogonal wavelets \cite{prasad2016performance} in the DWT structure to simulate LGN-V1 interactions. Both biorthogonal and Gabor wavelets are preferred wavelets for edge detection in images which are extensively employed to simulate the functioning of the V1 area in the primary visual cortex of primate vision systems \cite{fischer2007self, arai2005nonlinear, lee1996image}. To achieve this, the Large time-frequency analysis toolbox (LTFAT) \cite{pruuvsa2014large} and PyWavelets \cite{lee2019pywavelets, sondergaard2012linear} are used to apply DWT on raw images received at the retina to calculate edge-detected images as well as the reconstruction of transformed images.

In the current study, we investigated the impact of manipulating the number of low-frequency feature maps in the GFB on the extraction of higher spatial frequency content from input images. By adjusting the weights assigned to low-frequency filters, we were able to selectively control the emphasis on high-frequency components which results in edges of the original image. We apply this method on the 5, 10, 15, and 20 numbers of low-frequency feature maps named the GFB\_5, GFB\_10, GFB\_15, and GFB\_20 respectively, offering a customized and tailored feature extraction stage for our image processing task.

\subsection{Visual stimuli categorization with LGN activation}
\begin{figure}
	\centering
	\includegraphics[width=1\columnwidth]{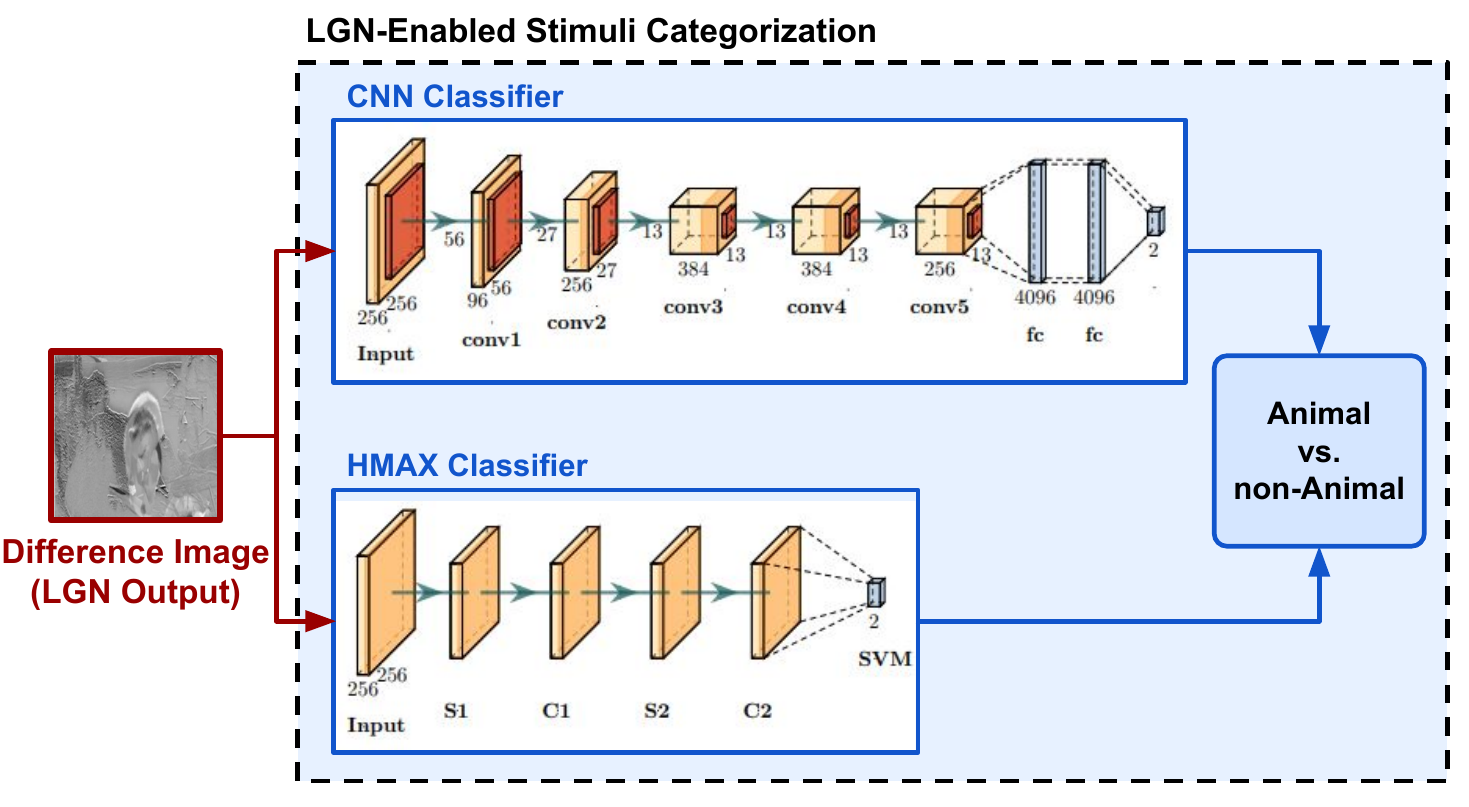}
	\caption{An illustration of the input visual categorization with the LGN area. This model uses the AlexNet (upper model) and also HMAX (lower model) classifiers to mimic the human ventral pathway. These two biologically inspired models are used to differentiate between animal and non-animal images which are initially preprocessed by the LGN area. Overall, this method offers an excellent framework to evaluate different models based on LGN preprocessed data.}
	\label{FIG:6}
\end{figure}

So far, we have explained the technique for modeling the LGN area using the proposed pruned convolutional autoencoder and also wavelet transformation. To evaluate the performance of the LGN model in categorizing input images, we aim to employ two approaches to present the ventral visual pathway as a general classifier: 1) a CNN method using the pre-trained AlexNet, and 2) the HMAX bio-inspired model. Both models mimic human hierarchical visual processing and also enable us to compare the categorization performance in both temporal (LGN-enabled) and non-temporal (without LGN) modes.
As depicted in \hyperref[FIG:6]{Fig. \ref{FIG:6}}, we used the AlexNet network as the first classifier due to its structural similarity to the human ventral visual pathway \cite{krizhevsky2012imagenet, hu2018cnn, zhuang2017deep}. AlexNet comprises a pre-designed network with eight weighted layers: the first five are convolutional layers, and the remaining three are fully connected layers. The output of the last fully connected layer is passed to a 1000-way softmax, producing a distribution over 1000 class labels. In this study, we adjusted the output size of the last fully connected layer to classify the input images into two object categories. We classify the data into animal and non-animal categories, and our model is expected to accurately distinguish these two classes. In addition to this, we utilized the HMAX model, a biologically inspired computational model that is designed to mimic visual information processing in the human visual cortex \cite{riesenhuber1999hierarchical}. It effectively recognizes various objects by modeling the visual system's processing in the V1, V4, and IT regions \cite{sufikarimi2020role}. The HMAX model generally consists of different layers, namely S1, C1, S2, and C2 layers, each of which carries out an individual operation to extract visual data. An image the retina receives passes through a network of cells and the HMAX model's S1 units are similar to the edge-detecting simple cells in the visual cortex. The S2 layer corresponds to cells in the V4 and IT layers. It calculates the difference for a single feature map across all positions and orientations within each scale band. The C2 layer response is determined by taking the global maximum across all scales and positions \cite{li2015enhanced}. Finally, we use a support vector machine (SVM) to differentiate between animal and non-animal data.

\subsection{Conducting the psychophysical experiment to validate models}
\begin{figure*}
	\centering
	\includegraphics[width=120mm]{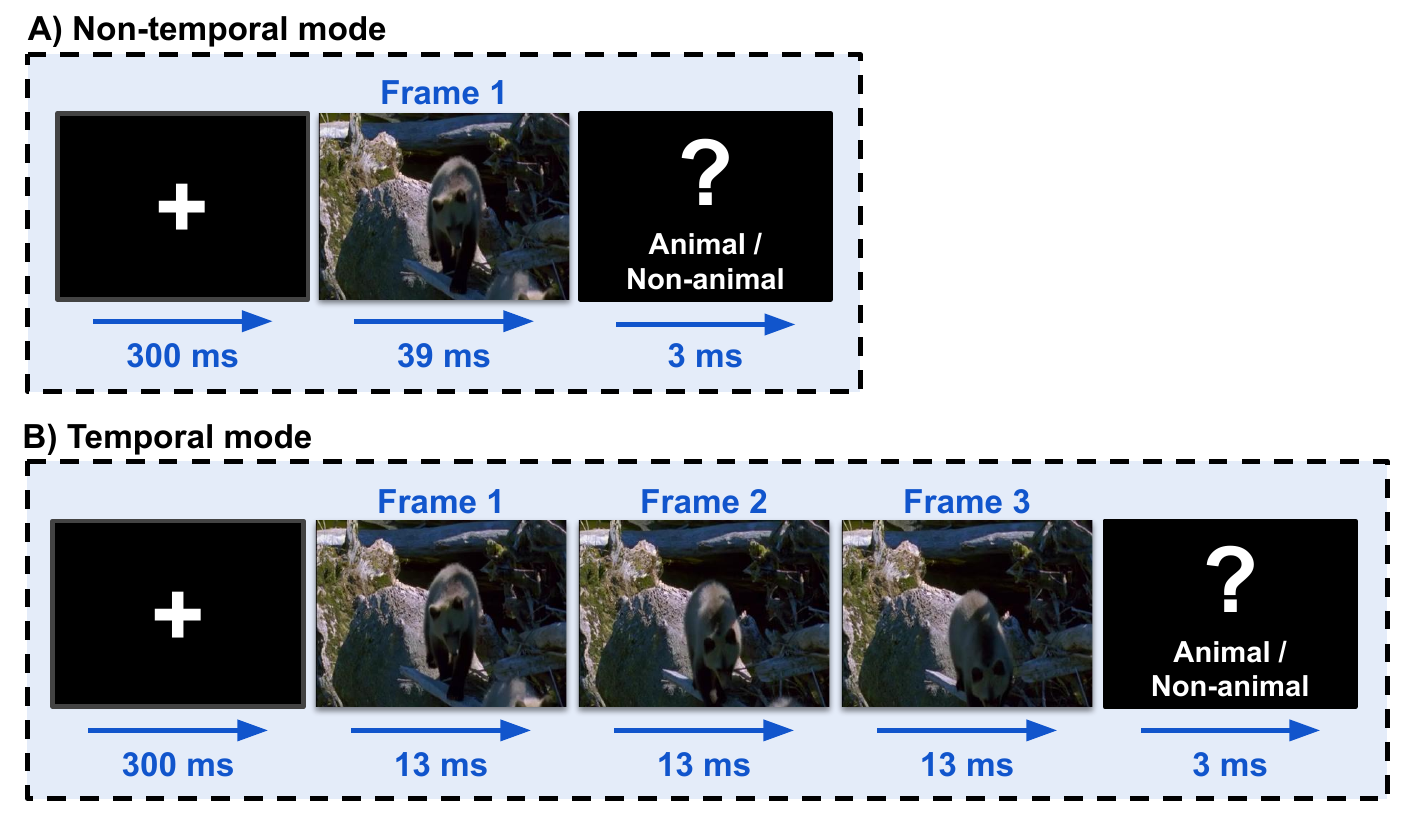}
	\caption{Illustration of the experimental setup for psychophysical test. (A) The non-temporal condition is where participants view single frames from a sequence. The sequence begins with a fixation cross displayed for 300 milliseconds, followed by a single frame shown for 39 milliseconds. After the frame presentation, participants had 3 milliseconds to respond, indicating whether they saw an animal or not. (B) The temporal condition where participants first see a fixation cross for 300 milliseconds, followed by three consecutive frames (each displayed for 13 milliseconds) showing an animal in motion. After the last frame, a screen prompts the participants to identify whether an animal was present or not within 3 milliseconds.}
	\label{FIG:7}
\end{figure*}

In this section, we aim to design an experiment to compare the results of our proposed models with human subjects. This experiment seeks to evaluate the effectiveness of the developed models with and without LGN compared to diverse people. We conducted experiments under two conditions: temporal and non-temporal image presentation. The participant pool also consisted of 8 females (aged 20-34 years) and 22 males (aged 22-33 years). All participants had normal eyesight and performed tests in a dark, noise-free room.
While the 80 random images are shown to participants, they are instructed to press one of two designated keys on a keyboard in a way that: if they see an animal in the image, they press the "yes" key, and if no animal is present, they press the "no" key. In the temporal condition, subjects observe 40 images in 3-frame sequences, while in the non-temporal condition, 40 images are shown as single frames. Each condition considers an equal number of frames for images with and without animals. The presentation time for each image is 39 milliseconds and participants have 3 seconds to respond for that particular frame. In the non-temporal condition, the single-frame presentation leads to the LGN area not being involved in decision-making, as illustrated in \hyperref[FIG:7]{Fig. \ref{FIG:7}}. As well, in the temporal mode, there are 40 sets of three-frame sequences from the movie, each illustrating scenes with dynamic changes, such as moving animals or humans. Each frame is displayed for 13 milliseconds, resulting in a total viewing duration of 39 milliseconds per sequence. We expect activation of the LGN in the brain due to continuous changes happening through the images in each sequence as depicted in \hyperref[FIG:7]{Fig. \ref{FIG:7}}. Importantly, to familiarize participants with the experiment and reduce task-induced effects, they performed similar tasks with different images as a set of training demos before the main experiment. Participants were tasked to indicate whether an animal was present in the image by pressing the "YES" or "NO" key on the keyboard.

\section{Results}
In this section, we first analyze the output results of the proposed models designed for the LGN with the GFB and the pAE processing models in feedforward (LGN-V1) and feedback (V1-LGN) streams. Next, we utilize these outputs in conjunction with two AlexNet and HMAX models, comparing the results to those obtained from experimental participants. We apply HMAX and CNN models in both temporal and non-temporal modes to compare results when data is processed in different directions. A psychological test also was administered using MATLAB \cite{brainard1997psychophysics} and PyTorch \cite{paszke2019pytorch} software to evaluate the proposed models compared to the actual human results. 

\subsection{Visualizing cross-sectional information stream in LGN-V1 connection}
In this section, we will examine the feedforward and feedback pathways in the LGN using different image processing methods i.e. the GFB and pAE models for edge detection, image reconstruction, and moving object detection tasks through each stream. We start with analyzing the feedforward stream from the retina to the LGN and then to V1, where edge detection is performed on the received image. Next, we explore the feedback stream from V1 to the LGN area to create more relevant visual information for further processing. Using inverse Gabor filter methods and also the pAE model, we reconstruct the original image from the V1 output. Finally, we use the reconstructed image along with the second input frame to detect variations in object recognition, specifically whether an animal is in the image or not.

\subsubsection{Exploring image edge-detection in V1 through forward flow of visual data}
\begin{figure*}
	\centering
	\includegraphics[width=170mm]{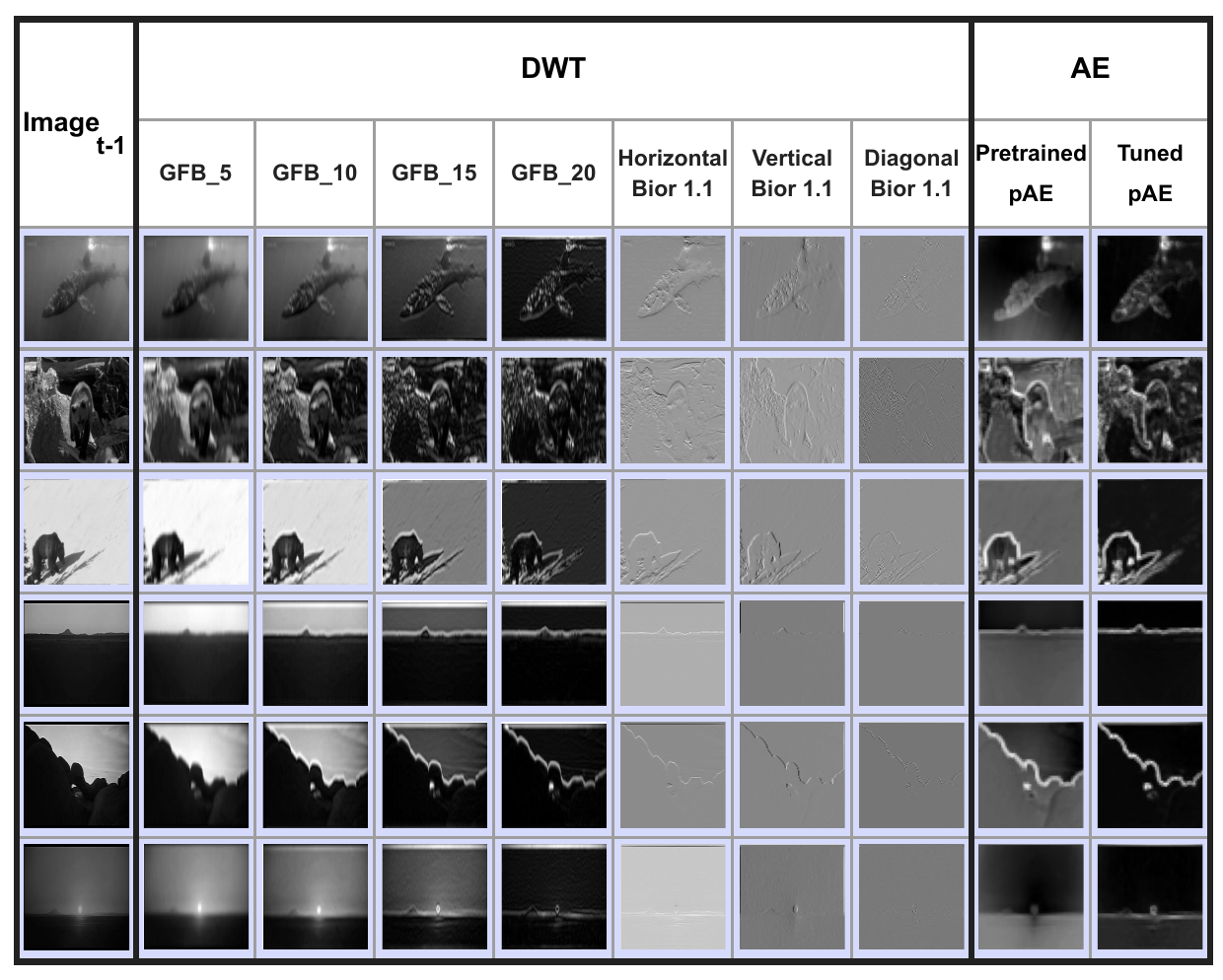}
	\caption{The illustration of edge-detected images in the V1 area. The input images from LGN to V1 are subjected to edge detection techniques, including DWT and AE, to provide fine information for further processing across the visual system. Note that, the first edge-detected images for GFB and pAE models over the desired frequency range are displayed due to the large number of sub-frequency images. Overall, the edge detection process allows for a more accurate interpretation of input images, contributing to understanding LGN functions through the visual system.}
	\label{FIG:8}
\end{figure*}

As mentioned earlier, the data transformation from LGN to V1 through the vision pathway represents edges in input images, reflecting the salient information received from visual stimuli.  The GFB, DWT, and pAE models are utilized to simulate this part of the vision pathway, where each method offers unique advantages in capturing a stimuli's fine details and characteristics. 
\hyperref[FIG:8]{Fig. \ref{FIG:8}} illustrates several images containing edges for animal and non-animal datasets at different setting parameters along the visual pathway. The GFB model produces 128 outputs, with four specific examples highlighted: GFB\_5, GFB\_10, GFB\_15, and GFB\_20. As depicted in this figure, the 2D detailed coefficients generated by the Gabor filtering method demonstrate that using various levels of detailed information enhances processing effectiveness and helps to distinguish the object from the irrelevant background. Similarly, the Tuned pAE model produces a more understandable output object compared to other methods. Among the different GFB models, GFB\_20 reflects more detailed information compared to the GFB\_5, GFB\_10, and GFB\_15, which is comparable to the tuned pAE model.

\subsubsection{Reconstructing images in LGN explains the backward flow of visual data}
\begin{figure}
	\centering
	\includegraphics[width=1\columnwidth]{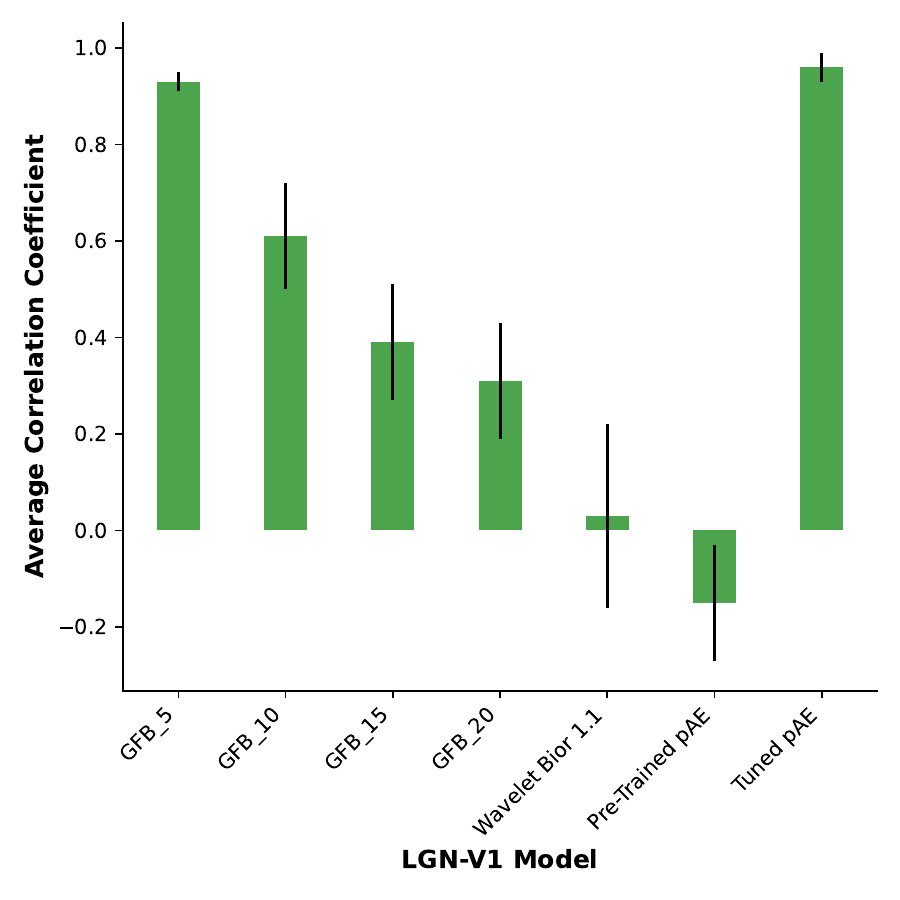}
	\caption{The reconstruction results for the feedback visual stream from the V1 to LGN. This process involves taking the edge-detected from V1 and applying various reconstruction methods, including the pre-trained pAE, tuned pAE, and DWT methods. The figure displays the average correlation coefficients between the original image and the LGN reconstructed image with corresponding standard deviation. Overall, the best-performing model generates reconstructed images with a refined and comprehensive representation of the original visual stimuli.}
	\label{FIG:9}
\end{figure}

To enhance the accuracy and clarity of the perceived image and refine visual information, the feedback mechanism from V1 to the LGN plays a crucial role in the human visual cortex. Thus, we now focus on reconstructing the image from the feedback output path from V1 to the LGN using the inverse Gabor filters and decoding part of the pAE model. The reconstructed images are generated based on the output from V1, which primarily consists of the edges detected in the input retina image.
We have computed the correlation coefficient between the original and reconstructed images to evaluate the reconstruction performance in the LGN-V1 connection. \hyperref[FIG:9]{Fig. \ref{FIG:9}} explains how each cortical processing method performs well in reconstructing information when passing data on the feedback stream. As this figure reveals, the Tuned pAE method achieves the highest reconstruction performance, while the GFB\_20 method exhibits poor reconstruction capabilities across all methods (0.960.03 and 0.030.19 for Tuned pAE and Bior 1.1 models, respectively). The variations in correlation values across different methods indicate significant differences in their effectiveness at reconstructing images. We have also provided some examples of the reconstructed images and corresponding performance in the LGN area for different input images (see supplementary material Fig. A.1). Notably, the presence of negative correlation values suggests that certain methods may introduce distortions in the reconstructed images. Overall, this comparative analysis provides valuable insights into the effectiveness of different feedback-based image reconstruction techniques, contributing to a broader understanding of visual processing in the brain.

\subsubsection{Detecting moving objects as the preferred output of the LGN area}
\begin{figure*}
	\centering
	\includegraphics[width=130mm]{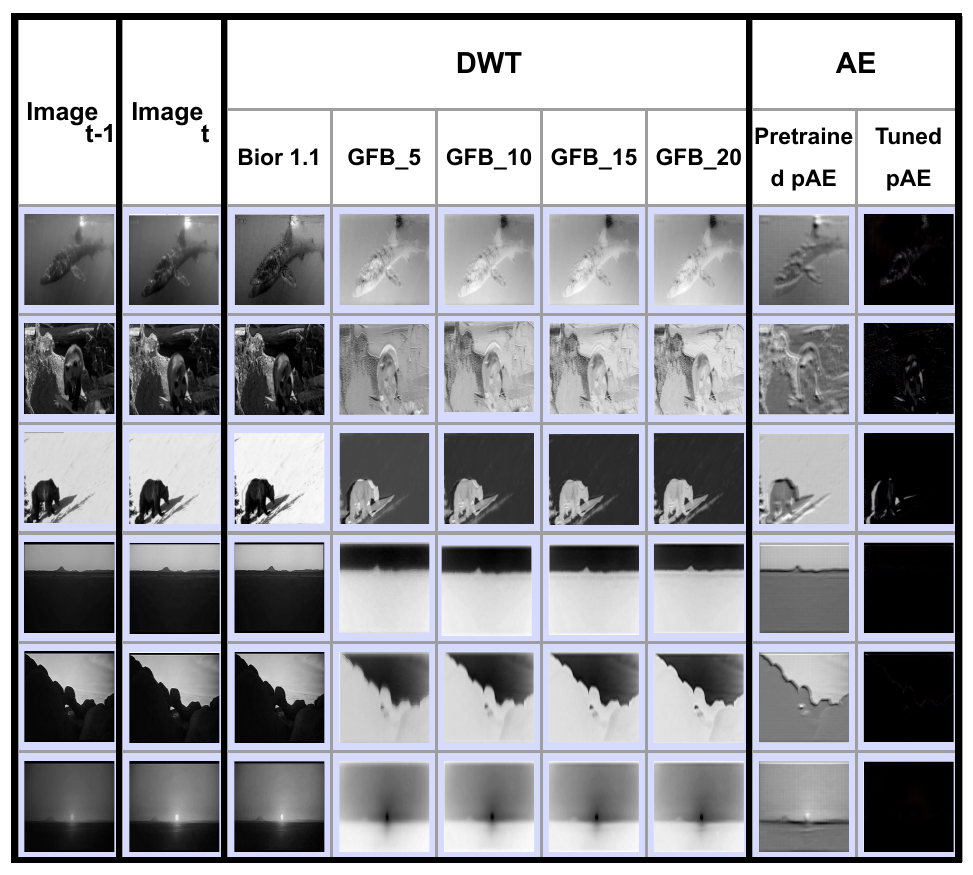}
	\caption{The temporal changes of the visual data stream in the LGN area. The output data of the LGN is created by computing the difference between the reconstructed image from the V1 feedforward flow to the LGN and the subsequent image received at the retina. Overall, this mechanism allows the visual system to adapt and respond efficiently to the alterations in visual stimuli. }
	\label{FIG:10}
\end{figure*}

In this stage, the difference between the reconstructed images (transferred back from V1 to LGN) and the second input image (transferred from the retina to LGN) is calculated to detect changes in the final processed image. The resulting image is then prepared for the subsequent classification stage. 
\hyperref[FIG:10]{Fig. \ref{FIG:10}} illustrates the final LGN output image derived from the second input frame and the reconstructed image computed across different methods. As we desired, this figure reveals in the category of animal images, where the object (i.e. animals) is present, moving, and sequentially fed into the visual system, we would expect to track the object dynamics in the output images of the LGN. The tuned pAE model successfully captured these changes, distinguishing it from other models that failed to properly and accurately identify the target object. Conversely, in the category of non-animal images, where no objects are present, we expected the output to consist of completely black images, reflecting the absence of object changes. The tuned pAE model depicted this result exactly as predicted. Also, all DWT-based models display a range of various 2D representations at different gray-scale levels of the original input image. This is exactly what occurs with the pre-trained pAE model.

\subsection{Evaluating the performance of object categorization models}
\begin{figure*}
	\centering
	\includegraphics[width=160mm]{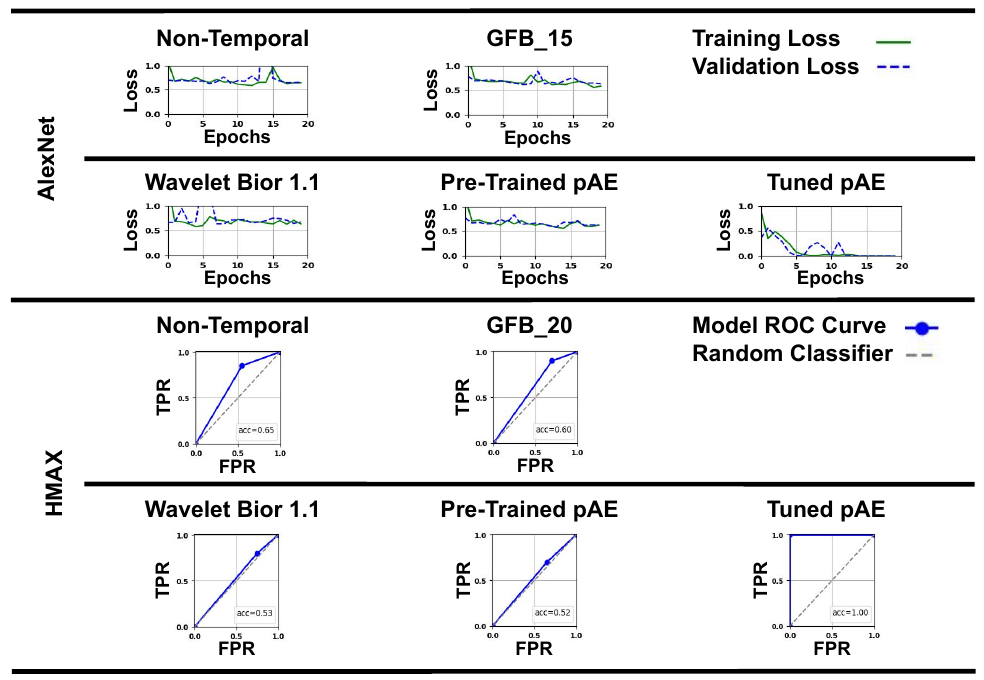}
	\caption{The training and validation performance for AlexNet and HMAX categorization models. The performance curves present the loss values for various models across consecutive epochs. Notably, the tuned pAE model demonstrates a lower loss than the other models.}
	\label{FIG:11}
\end{figure*}

We evaluated the performance of our proposed models, i.e. DWT and the AE-based models, by comparing their classification accuracy for animal and non-animal images under both temporal and non-temporal conditions. As mentioned earlier, we here employ the CNN and HMAX architectures to observe the prediction performance on these LGN-processed images. As a performance assessment, we compare the model-produced results with actual human object prediction performance in both temporal and non-temporal modes by computing the evaluation metrics including accuracy, precision, and recall. \hyperref[FIG:11]{Fig. \ref{FIG:11}} displays the training and validation performances for AlexNet and HMAX models across various LGN-V1 processing models.

{\hyperref[tbl1]{Table \ref{tbl1}}} presents the pairwise comparisons of different model configurations using the HMAX and AlexNet architectures. For each comparison, the table reports the difference in means, along with the corresponding 95\% confidence intervals (CI) and p-values. Several comparisons reveal statistically significant differences, particularly between GFB\_20 and Tuned pAE (the mean difference of -55.21 and a mean difference of -48.40 with a p-value of 0.00 for HMAX and AlexNet architectures, respectively) indicating a strong difference in performance. Conversely, the comparisons between GFB\_20, Bior 1.1, and Pre-Trained pAE show no significant difference, as reflected by a p-value of 1.00 for both HMAX and AlexNet models. Overall, this table underscores the varying degrees of performance differences between model configurations across the HMAX and AlexNet architectures, with certain model pairs showing significant improvements while others remain statistically indistinguishable.

\renewcommand{\arraystretch}{1.2}
\begin{table*}[h!]
\caption{Post-hoc test results for pairwise comparisons of HMAX and AlexNet categorization models}\label{tbl1}
\centering
\begin{tabularx}{\textwidth}{XXXXXXXXXX}
\hline
\multirow{2}{*}{} & \multicolumn{4}{c}{\hspace{100pt}HMAX} & \multicolumn{4}{c}{\hspace{100pt}AlexNet} \\ \cline{1-10}

\multicolumn{2}{c}{\centering Compared models} & \centering Lower limit for 95\% CI & \centering Difference of means & \centering Upper limit for 95\% CI & \centering p-value & \centering Lower limit for 95\% CI & \centering Difference of means & \centering Upper limit for 95\% CI & p-value \\ \hline

\centering GFB\_20 & \centering Bior 1.1 & \centering -7.96 & \centering 0.25 & \centering 8.46 & \centering 1.00 & \centering -10.35 & \centering -0.24 & \centering 10.84 & 1.00 \\ 

\centering GFB\_20 & \centering Pre-Trained pAE & \centering -7.21 & \centering 1.00 & \centering 9.21 & \centering 1.00 & \centering -14.50 & \centering -3.90 & \centering 6.7 & 1.00 \\ 
\centering GFB\_20 & \centering Tuned pAE & \centering -55.21 & \centering -47.00 & \centering -38.79 & \centering 0.00 & \centering -48.40 & \centering -37.80 & \centering -27.2 & 0.00 \\ 
\centering GFB\_20 & \centering Non-Temporal & \centering -19.46 & \centering -11.25 & \centering -3.04 & \centering 0.01 & \centering -15.23 & \centering -4.63 & \centering 5.96 & 1.00 \\ 
\centering Bior 1.1 & \centering Pre-Trained pAE & \centering -7.46 & \centering 0.75 & \centering 8.96 & \centering 1.00 & \centering -14.74 & \centering -4.15 & \centering 6.45 & 1.00 \\ 
\centering Bior 1.1 & \centering Tuned pAE & \centering -55.46 & \centering -47.25 & \centering -39.04 & \centering 0.00 & \centering -48.64 & \centering -38.04 & \centering -27.44 & 0.00 \\ 
\centering Bior 1.1 & \centering Non-Temporal & \centering -19.71 & \centering -11.50 & \centering -3.29 & \centering 0.01 & \centering -15.48 & \centering -4.88 & \centering 5.72 & 1.00 \\ 
\centering Pre-Trained pAE & \centering Tuned pAE & \centering -56.21 & \centering -48.00 & \centering -39.79 & \centering 0.00 & \centering -44.50 & \centering -33.90 & \centering -23.3 & 0.00 \\ 
\centering Pre-Trained pAE & \centering Non-Temporal & \centering -20.46 & \centering -12.25 & \centering -4.04 & \centering 0.01 & \centering -11.33 & \centering -0.73 & \centering 9.87 & 1.00 \\ 
\centering Tuned pAE & \centering Non-Temporal & \centering 27.54 & \centering 35.75 & \centering 43.96 & \centering 0.00 & \centering 22.57 & \centering 33.16 & \centering 43.76 & 0.00 \\ \hline

\end{tabularx}
\end{table*}

\begin{figure*}
	\centering
	\includegraphics[width=140mm]{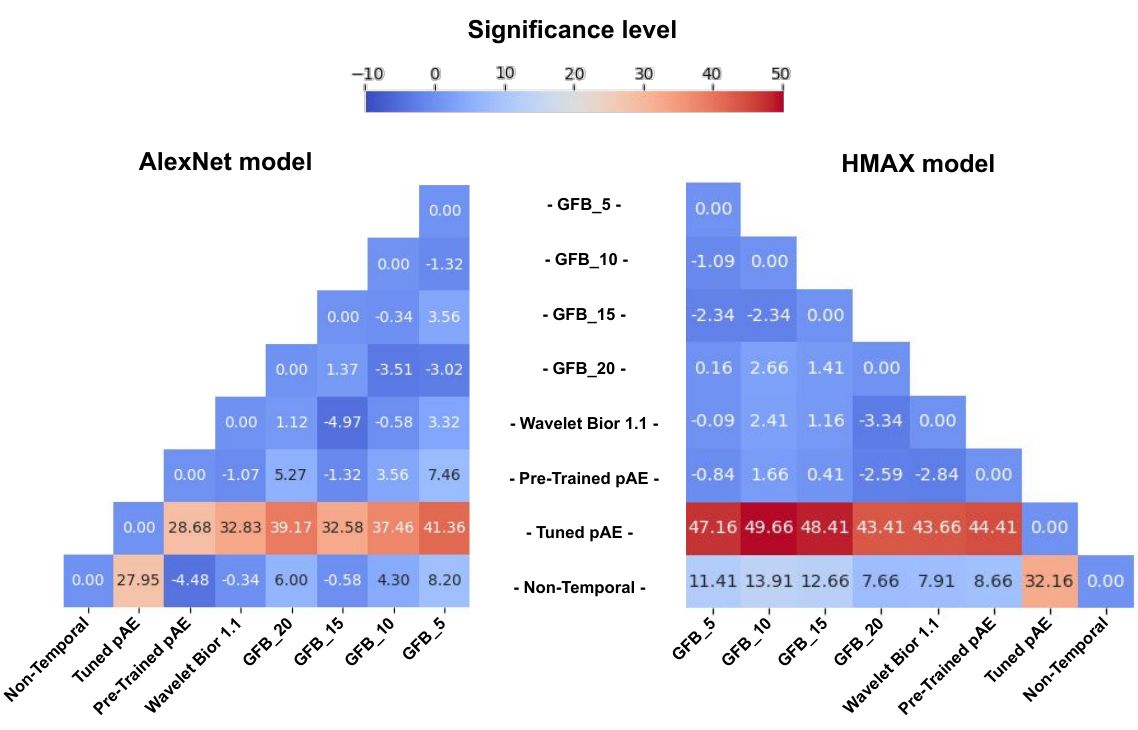}
	\caption{The heatmap for significance score between cortical processing methods within the HMAX and AlexNet architectures. The tuned pAE model achieves the higher significance level producing distinguishable results compared to other models when predicting true image labels.}
	\label{FIG:12}
\end{figure*}

As depicted in \hyperref[FIG:12]{Fig. \ref{FIG:12}}, we formally define the significance score to demonstrate how different cortical processing methods yield distinct results when compared to one another. Considering the performance results for two compared models, the significance score is defined as:
\begin{flalign}
    Significance \, score &= (m_{1} - SE_{1}) - (m_{2} + SE_{2}) &\nonumber\\ for  \,\, m_{1}>m_{2}\label{eq6}
\end{flalign}

Where the $m_1$ and $m_2$ denote the average prediction performance with corresponding standard errors $SE_1$ and $SE_2$. The positive significance scores present non-overlapping ranges of performance values revealing distinguishable results between comparing models. As depicted in this figure, the tuned pAE model achieves a higher significance score which is particularly evident when it comes to accurately predicting image labels. The ability of the tuned pAE model to generate distinct results suggests that the underlying architecture may be more effective in modeling the LGN-V1 connection and integrating feedforward and backward streams.

\subsection{Achievements for proposed models follow psychophysical human results}
\begin{figure}[htbp]
	\centering
	\includegraphics[width=1\columnwidth]{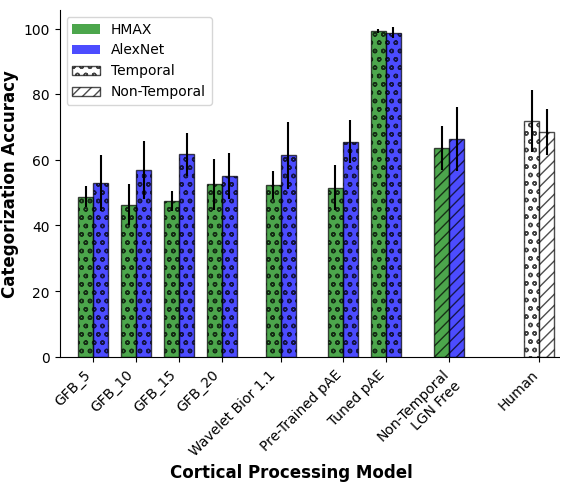}
	\caption{The accuracy percentage results for both temporal and non-temporal modes. This comparison is conducted using a diverse range of processed visual data such as DWT, pre-trained pAE, and tuned pAE to capture the effectiveness of the classification models in processing different types of data. Overall, the computational models including HMAX and pre-trained AlexNet attain appropriate performance in biological image processing compared to human outcomes.}
	\label{FIG:13}
\end{figure}

To run our behavioral test, we used the Psychophysics Toolbox implemented in MATLAB software to measure the human prediction performance across different subjects. These psychophysical experiments were conducted with 30 human subjects seated 0.5 meters from a computer screen (Intel Core i7 processor, 8 GB RAM) running MATLAB software with the Psychology Toolbox. As depicted in \hyperref[FIG:13]{Fig. \ref{FIG:13}}, the participants achieved an average accuracy of 67.50±6.95\% in the non-temporal mode, which increased to 72.50±9.48\% in the temporal mode. These findings suggest that the temporal processing mechanism significantly enhances visual perception and classification accuracy.

Besides, our approach yields significantly better results than other methods in the temporal mode, with an approximate improvement of 28\% over the actual human result when using the tuned pAE model coupled with HMAX and also the prediction improvement of 27\% when using the tuned pAE model coupled with the AlexNet classification model. On the contrary, the actual human subjects in the non-temporal mode outperformed the corresponding non-temporal LGN-free models by 2\% and 4\% for the AlexNet and HMAX models, respectively. In addition, to emphasize the importance of LGN modeling in object categorization, this figure reveals that using a tuned pAE model with both classification models significantly improves prediction performance compared to the LGN-free model. Overall, the tuned pAE demonstrates appropriate effectiveness in processing visual data and achieves great prediction performance in combination with the HMAX and pre-trained AlexNet, surpassing human outcomes in biological image processing.

\section{Discussion}
In the current study, the proposed tuned pAE model demonstrates superior performance than human results, indicating that it is a reliable tool for simulating human vision. We evaluated the model in both temporal and non-temporal modes to better understand the role of the LGN in visual processing. This dual-mode approach provides deeper intuitions into the role of LGN in the brain's visual functions. In the temporal mode, in which the LGN component was included, the model achieved better results than when this component was removed. LGN integration increases the model's ability to detect changes and movements in images. Feedback streams from V1 to the LGN allow the model to detect temporal and spatial changes more effectively, thereby improving its performance in detecting moving objects. As an important step in the visual pathway, the LGN processes and amplifies information in a way that significantly enhances the performance of visual processing models.
The tuned pAE model performed better than the wavelet model, primarily because the tuned pAE model, which is a deep neural network, can automatically learn complex features from the data. This allows it to effectively identify non-linear features and complex patterns in images. In contrast, the wavelet method, which decomposes the signal into different frequency components, is more suitable for analyzing simpler and more linear features. Therefore, in scenarios involving complex patterns or nonlinear relationships, the wavelet method may not perform as well as the tuned pAE model. The tuned pAE model's deep neural network structure also enables it to reconstruct images more accurately, as it can learn and retain essential information from the images. On the other hand, the wavelet method, which is commonly used for signal compression and decomposition, may have difficulty in accurately reconstructing images and preserving all details. Furthermore, the tuned pAE model outperformed human participants. It can be said that due to the ability to identify key features that are critical to detecting patterns that may not be obvious to humans, this advantage allows the model to achieve greater accuracy in certain scenarios. Unlike humans, the tuned pAE model is not affected by psychological factors, fatigue, or distraction, ensuring stable and reliable performance, while human accuracy can fluctuate under different conditions.

\subsection{Novelty and contribution of the current study}
The novelties and contributions presented in this study are outlined in detail as follows.

\begin{itemize}[itemindent=0em,leftmargin=1.5em]

\item \textbf{Introduction of the Tuned pAE model for LGN simulation:} The tuned pAE model specifically was designed to simulate various functions within the LGN region. This customized single-layer convolutional encoder-decoder was proposed to approximate both feedforward and feedback interactions between the LGN and V1 areas.
\item \textbf{Enhanced temporal mode performance:} The inclusion of the LGN component in the temporal mode significantly improved the model's performance compared to when the component was excluded.
\item \textbf{Investigation into interpretable CNN-based pAE model compared to classical models:} The study included the design of various filtering models using GFB and DWT for a comprehensive analysis of neural pathways involved in visual perception. However, the tuned pAE model, being a deep neural network, demonstrated better performance by effectively learning complex features and non-linear patterns, and achieving more accurate image reconstruction. 

\item \textbf{Biologically plausible visual processing pipeline:} The study focused on modeling the LGN and simulating the ventral pathway with hierarchical classification models, incorporating a feedback mechanism from V1 to LGN to develop a biologically plausible visual processing pipeline.

\item \textbf{Achieving superior performance in classification tasks:} The tuned pAE model outperformed other models, particularly when combined with AlexNet, in classification tasks. The outcomes of the model were evaluated against the widely recognized HMAX model and validated through a real psychophysical experiment involving 30 participants.

\end{itemize}

\subsection{Limitations}
One primary limitation is the model's dependence on certain assumptions and simplifications in replicating the complex mechanisms of the human visual system. For instance, although the model incorporates both feedforward and feedback streams, it may not fully capture the intricate neural interactions and dynamic processes that occur in the brain, particularly in response to more complex visual stimuli or tasks beyond object recognition. Furthermore, the psychophysical experiments conducted to compare the model's performance with human vision were limited in scope, involving a relatively small sample size of participants and specific experimental conditions. This may limit the generalizability of the results to wider populations or different experimental settings.

\subsection{Future research areas}
Our dual-mode approach, which includes both parts of the LGN and the visual cortex, provides deeper insight into the brain's visual processing mechanisms. Future work will focus on further refining the model to increase its accuracy and extend its applicability to more complex visual tasks. We also plan to explore the integration of additional neural components and feedback mechanisms from higher layers to create a more accurate representation of the human visual system.
Moreover, employing more novel categorization architectures that integrate different connections including lateral, and recurrent into the model, will be an appropriate idea for model enhancement. It is also possible to extend the study to more complex visual tasks, such as distinguishing between objects such as humans, cats, dogs, and birds. This extension helps validate the model's adaptability and effectiveness across various visual challenges.

\section{Conclusion}
In this study, we developed a comprehensive framework that mimics the human visual system in both temporal and non-temporal modes of operation. Our framework uses a novel modeling method to simulate the LGN function in feedforward and feedback pathways. This model incorporates the data stream from LGN to V1 by approximating the image edges and conversely by reconstructing the processed data in reverse stream from V1 to LGN, generating the actual LGN output for further analysis in downstream areas. Next, the LGN-processed image is then classified by the AlexNet and HMAX classification models. We primarily aim to develop a detailed architecture that accurately replicates the interactions at the LGN-V1 connection, including both feedforward and feedback data flows. Our findings revealed that the proposed models closely mimic human visual processing, with the tuned pAE model significantly outperforming the other models in terms of prediction performance.

Our findings also highlight the importance of the role of the LGN in visual processing, especially during the feedback flow of visual data, which increases the clarity and accuracy of reconstructed images. Finally, the proposed model holds promise as an efficient tool for simulating human visual processing, contributing to advances in human vision research.

\appendix
\section{Appendix}
\textcolor{white}{a}

\renewcommand{\thefigure}{A.\arabic{figure}}
\setcounter{figure}{0}

\begin{figure*}[H]
	\centering
	\includegraphics[width=140mm]{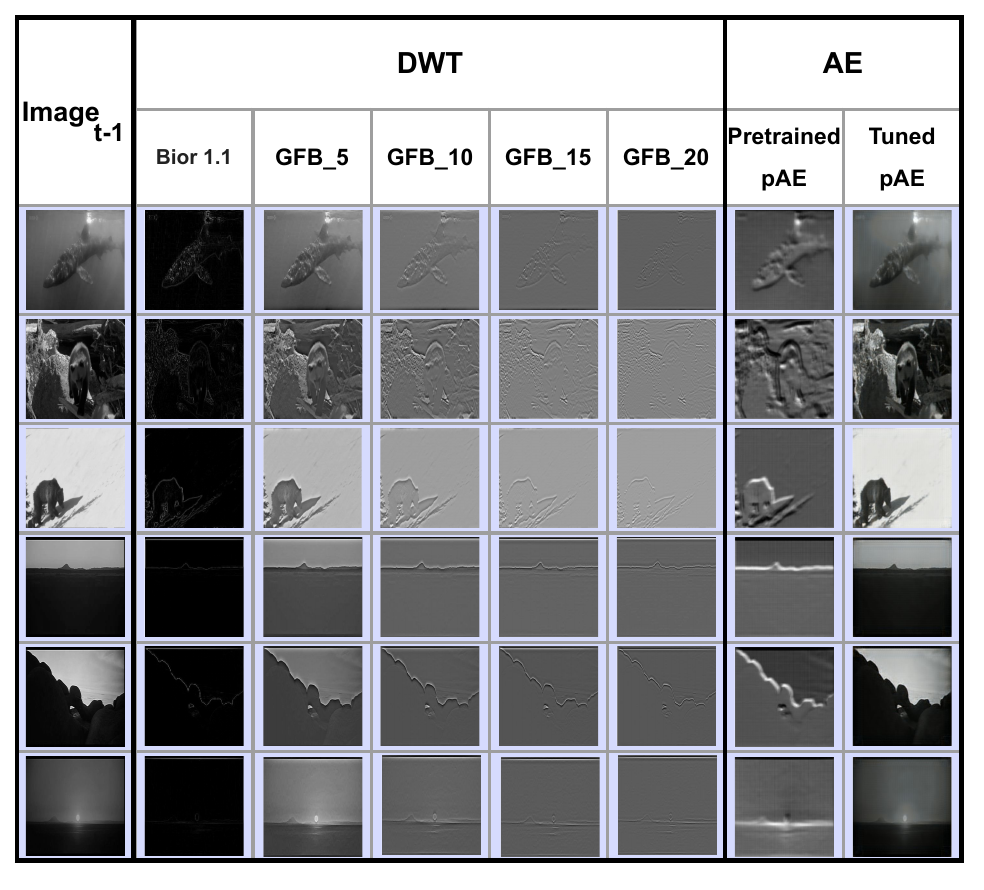}
	\caption{The sample reconstructed images by different processing models for the LGN-V1 connection.}
	\label{FIG:A1}
\end{figure*}

\printcredits

\bibliographystyle{cas-model2-names}

\bibliography{main}

\end{document}